# Non-sensitive axis feedback control of test mass in full-maglev vertical superconducting gravity instruments

Daiyong Chen[1], Xikai Liu[2,*], Lulu Wang[1], Liang Chen[1] and Xiangdong Liu[1,†]

[1]*MOE Key Laboratory of Fundamental Physical Quantities Measurement & Hubei Key Laboratory of Gravitation and Quantum Physics, PGMF and School of Physics, Huazhong University of Science and Technology, Wuhan 430074,  P. R. China*

[2]*Ningbo Institute of Technology, Beihang University ( BUAA), Ningbo, 315800,  P. R. China.*

Non-sensitive axis feedback control is crucial for cross-coupling noise suppression in the application of full-maglev vertical superconducting gravity instruments. This paper introduces the non-sensitive axis feedback control of the test mass in a home-made full-maglev vertical superconducting accelerometer. In the feedback system, special superconducting circuits are designed to decouple and detect the multi-degrees-of-freedom motions of the test mass. Then the decoupled motion signals are dealt with by the PID controller and fed back to the side-wall coils to control the test mass. In our test, the test mass is controlled successfully and the displacement is reduced by about one order of magnitude in the laboratory. Accordingly, the noise level of the vertical superconducting accelerometer in the sensitive axis is also reduced.

[*]liuxikai@hust.edu.cn

[†]liuxd@hust.edu.cn

# I. INTRODUCTION

In full-maglev vertical superconducting gravity instruments (VSGI) [1,3], non-mechanical connection exists between the superconducting test mass and the base. In order to keep the test mass levitated stably, side-wall superconducting-current-carrying-coils [2] must be used to provide the needed stiffness to suppress the motion of the test mass in the non-sensitive axis, thus reducing the cross-coupling noise [5,6]. Due to the limit of the critical magnetic field of superconductor, the current in the side-wall coils cannot be too large, so the stiffness provided by these coils is also limited. In the quiet laboratory, the stiffness may be large enough to achieve a good cross-coupling noise reduction. However, in the dynamic environment, such as the field outside the laboratory, the quiet and even the plane [9-11], the vibration level is many orders of magnitude higher than in the laboratory and the cross-coupling noise will be much larger. In order to reduce the cross-coupling noise in the dynamic noise, one needs to further suppress the motion of the test mass in non-sensitive axis in case that the current in the side-wall coils cannot be increased. The feedback control is the most suitable technique. By the non-sensitive axis feedback control, the equivalent stiffness can be improved significantly.

Up to now, few papers about the feedback control of the test mass in full-maglev superconducting gravity instruments have been reported, owing to the complex coupling relationship in the system. The motion of the full-maglev test mass has multi-degrees-of-freedom [12]. Due to the parameter mismatch and installation asymmetry of side-wall coils, there will be different types of coupling in the system, such as magnetic coupling between side-wall coils and motion coupling in different axes. The key for the control lies in the decoupling of the system. One method has been utilized in Ref. [15], in the method, superconducting circuits constructed with 8 side-wall coils at low temperature and gain adjustable amplifiers at room temperature are combined to determine the difference between the scale factors of the translational and rotational mode, and then motion of the test mass can be decoupled by data processing. This method does not introduce additional noise when the VSGI works in the open-loop state. While in the closed-loop state, the designed controller is much more complex compared with the PID controller [13,14], additional negative effects, such as the noise and time delay of the data processing, will be introduced into the system.

To simplify the control, this paper proposed another motion decoupling method. In our method, 16 side-wall coils were used to design special translational and rotational superconducting detecting circuits. After precise attitude adjustment of the test mass, these superconducting circuits can decouple the motion of the test mass directly. Then the non-sensitive axis feedback control of the test mass can be carried out with the PID controller. This method has been verified in the new-made VSGI. The motion of the test mass was decoupled and detected by the superconducting circuits, the output signals of the circuits were then dealt with by PID controller and feedback to the side-wall coils. The motion of the test mass in the non-sensitive axes was controlled successfully. The equivalent stiffness was increased by more than one order of magnitude. As a consequence, the noise level of the detected signal in the sensitive axis was also reduced.

# II. PRICIPLE AND METHEOD

## A. Superconducting decoupling circuits design

The working principle of the VSGI has been described in detail in Ref. [3-7]. In the sensitive axis, planar coils are used to levitate the test mass to construct the magnetic spring-oscillator structure, the displacement of the test mass is detected by the superconducting circuits and superconducting quantum interference device (SQUID) [16-18]. In the non-sensitive axis, side-wall coils are used to provide the needed stiffness to suppress the motion of the test mass. The structure of the new-made VSGI is similar to the traditional ones except for the quantity and distribution of the side-wall coils. 16 side-wall coils with the same designed electromagnetic and geometrical parameters are evenly mounted on the inner wall of the supporting cylinder, as shown in FIG. 1.

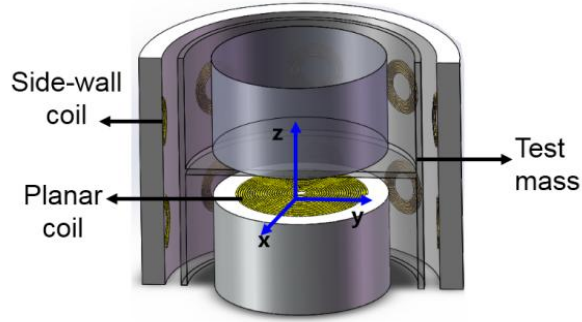

FIG. 1. Basic structure of the new-made VSGI

The side-wall coils are divided into four groups, each group has four designated side-wall coils. These four coils are connected to form the translational and rotational superconducting detecting circuits. The structure of the circuits is shown in FIG. 2. FIG. 2(a) is the translational detecting circuit, side-wall coils on the same side of the test mass are connected in series. FIG. 2(b) is the rotational detecting circuit, diagonal side-wall coils on two sides of the test mass are connected in series. The superconducting transformer and the SQUID on the right side detect the displacement of the test mass in this axis, and the superconducting transformer on the left side couples the feedback current into the circuit to generate the feedback forces for controlling the test mass.

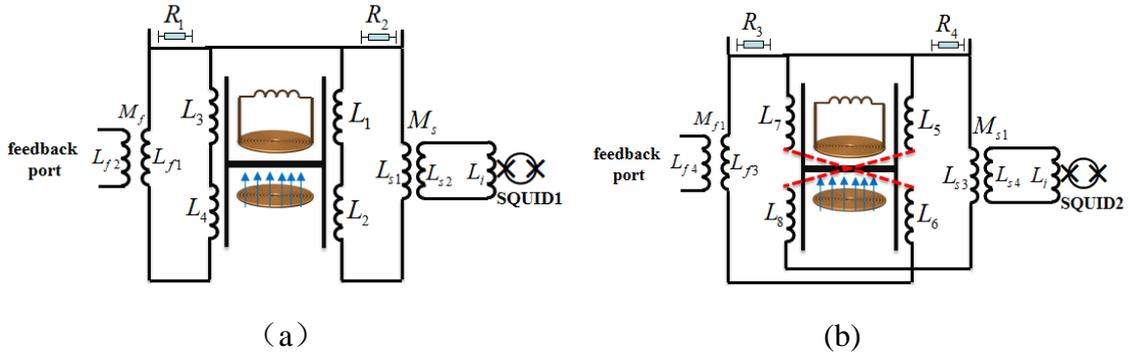

(a)          (b)

FIG. 2. (a) translational detecting circuit and (b) rotational detecting circuit.

Detailed calculations of the translational detecting circuit are given here to explain the working principle, and the rotational circuit is similar. In order to be close to reality, mismatch of the side-wall coils is taken into account. After the test mass is levitated, persistent current $I_0$ is injected into the side-wall coils to keep the mass at equilibrium position. Supposing the test mass has tiny translational displacement $x$ towards the right and rotational angle $\theta$ towards the left, the inductance of the four side-wall coils can be approximately written as [1,7]:

$$\begin{cases} L_1 = L_{10}(1+\lambda_1(-x+l\theta)) \\ L_2 = L_{20}(1+\lambda_2(-x-l\theta)) \\ L_3 = L_{30}(1+\lambda_3(x-l\theta)) \\ L_4 = L_{40}(1+\lambda_4(x+l\theta)) \end{cases} \quad (1)$$

Here, $L_{i0}$ (i=1...4) is the inductance of the four coils when the test mass is at the equilibrium position; $\lambda_i$ (i=1...4) is the linear coefficient associated with the displacement; $l$ is the length of the rotation arm.

The flux conservation of the superconducting circuits in FIG. 2 gives:

$$\begin{cases} (L_1 + L_2 + L_e)I_{t1} = (L_{10} + L_{20} + L_e)I_0 \\ (L_3 + L_4 + L_{f1})I_{t2} = (L_{30} + L_{40} + L_{f1})I_0 \end{cases} \quad (2)$$

Here

$$L_e = L_{s1} - \frac{M_s^2}{L_{s2} + L_i} \quad (3)$$

is the equivalent inductance of the detecting transformer load; $L_{s1}$, $L_{s2}$ and $M_s$ are the primary, secondary and mutual-inductance respectively; $I_0$ is the persistent current injected into the circuit; $I_{t1}$ and $I_{t2}$ are the changing current related to the displacement of the test mass; $L_{f1}$ is the primary inductance of the feedback transformer.

Combining equation (1) and (2), we can get the relationship of $I_{t1}$, $I_{t2}$ and the displacement of the test mass. After Taylor series expansion, the result is:

$$\begin{cases} I_{t1} = (1 + Ax - Bl\theta)I_0 \\ I_{t2} = (1 + Cx - Dl\theta)I_0 \end{cases} \quad (4)$$

Here

$$\begin{cases} A = \frac{L_{10}\lambda_1 + L_{20}\lambda_2}{L_{10} + L_{20} + L_e}, & B = \frac{L_{10}\lambda_1 - L_{20}\lambda_2}{L_{10} + L_{20} + L_e} \\ C = \frac{L_{30}\lambda_3 + L_{40}\lambda_4}{L_{30} + L_{40} + L_{f1}}, & D = \frac{L_{30}\lambda_3 - L_{40}\lambda_4}{L_{30} + L_{40} + L_{f1}} \end{cases} \quad (5)$$

If $L_{10}\lambda_1 = L_{20}\lambda_2$, and $L_{30}\lambda_3 = L_{40}\lambda_4$ could be achieved, then B and D will be zero, and equation (4) can be simplified to:

$$\begin{cases} I_{t1} = (1 + Ax)I_0 \\ I_{t2} = (1 + Cx)I_0 \end{cases} \quad (6)$$

In this case, the current detected by the superconducting circuit and the SQUID is proportional to the translational displacement of the test mass. So as for the rotational detecting circuit, the final result is:

$$\begin{cases} I_{r1} = (1 + El\theta)I_0 \\ I_{r2} = (1 + Fl\theta)I_0 \end{cases} \quad (7)$$

Here

$$\begin{cases} E = \frac{L_{50}\lambda_5 + L_{80}\lambda_8}{L_{50} + L_{80} + L_{e1}}, & F = \frac{L_{60}\lambda_6 + L_{70}\lambda_7}{L_{60} + L_{70} + L_{f3}} \end{cases} \quad (8)$$

and $L_{i0}$ (i=5...8) is the inductance of the four coils in FIG. 2 (b) when the test mass is at the equilibrium position; $\lambda_i$ (i=5...8) is the linear coefficient associated with the displacement.

The specially designed superconducting circuits in FIG. 2 can decouple the translation and rotation of the test mass in case that the parameters of the side-wall coils should be matched. Though the side-wall coils were precisely matched with the same designed electromagnetic and geometrical parameters, due to the installation error, the test mass is always in tilt when levitated. The tile will introduce mismatch for the parameters of the side-wall coils. So the key to eliminating the mismatch is adjusting the attitude of the test mass, ensuring that the test mass and the side-wall coil base are coaxial. In practical, one could use the cross-coupling coefficients between the non-sensitive and sensitive axes as the criterion for the adjustment (see section Ⅲ.A).

## B. Control model of the test mass displacement

Once the motion of the test mass in non-sensitive axis is decoupled by the superconducting circuits in FIG. 2, the coupled four-degrees-of-freedom motion system can be simplified as the combination of four independent one-degree-of-freedom systems. And the control model of the one-degree-of-freedom system is shown in FIG. 3:

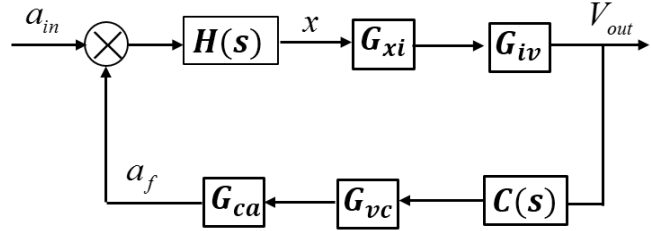

FIG. 3. one-degree-of-freedom control model

Here, $a_{in}$ and $a_f$ are the input and feedback accelerations respectively. $H(s)$ is the transfer function of the translational or rotational spring oscillator, which transfers the acceleration signal to the displacement of the test mass. For translation or rotation mode, the transfer function can be written as

$$H(s) = \frac{1}{s^2 + 2\xi\omega_n s + \omega_n^2} \quad (9)$$

$\omega_n$ and $\xi$ are the natural frequency and the damping ratio of the spring oscillator. $\omega_n$ is determined by the superconducting circuits and can be calculated according to the method in Ref. [5,20]. For translation and rotation mode, the natural frequencies are

$$\begin{cases} \omega_{nt}^2 = \frac{4}{m}\left(\frac{L_{01}^2 \lambda^2 I_0^2}{2L_{01} + L_t} + \frac{L_{01}^2 \lambda^2 I_0^2}{2L_{01} + L_{f1}}\right) \\ \omega_{nr}^2 = \frac{4l^2}{I}\left(\frac{L_{01}^2 \lambda^2 I_0^2}{2L_{01} + L_t} + \frac{L_{01}^2 \lambda^2 I_0^2}{2L_{01} + L_{f1}}\right) \end{cases} \quad (10)$$

The damping mainly comes from the residual gas in the vacuum where the VSGI works.

$G_{xi}$ is the scale factor between the displacement of the test mass and the current detected by the SQUID input coil, and can be obtained from equation (6) and (7):

$$\begin{cases} G_{xi-t} = \frac{M_s}{L_{s2} + L_i} AI_0 \\ G_{xi-r} = \frac{M_{s1}}{L_{s4} + L_i} EI_0 \end{cases} \quad (11)$$

$G_{iv}$ is the scale factor of the current to voltage conversion by the SQUID, and is $10^6$ when SQUID works in the medium sensitivity mode. C(s) is the transfer function of the controller. Here the practical PID controller is used. The transfer function of PID controller is in parallel form:

$$C(s) = P + \frac{I}{s} + Ds \quad (12)$$

$G_{vc}$ is the scale factor of the voltage-to-current converter. $G_{ca}$ is the scale factor of the feedback current to the feedback acceleration.

In the system, side-wall coils $L_3$ and $L_4$ work as the translational feedback acceleration actuator, $L_6$ and $L_7$ as the rotational one. When parameters of the side-wall coils are the same, the forces produced by the feedback current if injected into the feedback coils can be written as:

$$\begin{cases} F_3 = F_4 = \frac{1}{2}L_0\lambda(I_0 + i_{ft})^2 \approx \frac{1}{2}L_0\lambda I_0^2 + L_0\lambda I_0 i_{ft} \\ F_7 = \frac{1}{2}L_0\lambda(I_0 + i_{fr})^2 \approx \frac{1}{2}L_0\lambda I_0^2 + L_0\lambda I_0 i_{fr} \\ F_6 = -\frac{1}{2}L_0\lambda(I_0 + i_{fr})^2 \approx -\frac{1}{2}L_0\lambda I_0^2 - L_0\lambda I_0 i_{fr} \end{cases} \quad (13)$$

Then the translational feedback acceleration and the rotation feedback moment are calculated to be:

$$\begin{cases} a_{ft} = \frac{F_3 + F_4}{m} = \frac{1}{m}\left(L_0\lambda I_0^2 + 2L_0\lambda I_0 i_{ft}\right) \\ a_{fr} = \frac{(F_7 - F_6)\cdot l}{I} = \frac{1}{I}\left(lL_0\lambda I_0^2 + 2lL_0\lambda I_0 i_{fr}\right) \end{cases} \quad (14)$$

The first items on the right side of equation (14) are the static force and moment, equal to the force and moment produced by the rest coils in FIG. 2. The second items are proportional to the translational and rotational feedback current, thus controlling the motion of the test mass.

Once the parameters of the components in the system are known, the transfer functions and scale factors of each link will be determined and the control model can be simulated by the Simulink software to give the proper parameters of the PID controller.

The practical parameters of the components in the system are summarized in TABLE I.

TABLE I. related design parameters

| Parameter | Symbol | Value |
|---|---|---|
| **Mechanical** | | |
| test mass | $m$ | 0.1kg |
| moment of inertia | $I$ | $4.4\times10^{-4}$ kg m$^2$ |
| force arm | $l$ | 0.0125m |
| **Side-wall Coli** | | |
| Inner diameter | | 6mm |
| Outer diameter | | 18mm |
| Number of turns | | 45 |
| Spacing to test mass | | 0.5mm |
| Inductance | $L_{i0}$ | 6.5 μH |
| 1st derivatives of coils inductance | $\lambda L_{i0}$ | 0.01H/m |

| **Inductance of transformers (μH)** | | | | | | |
|---|---|---|---|---|---|---|
| $L_{s1}$ | $L_{s2}$ | $M_s$ | $L_{f1}$ | $L_{f2}$ | $M_f$ | $L_i$ |
| 1.6 | 35.4 | 0.3 | 1.6 | 33.1 | 3 | 2.6 |

In the simulation, the bandwidth and the phase margin are set as 30Hz and 60° respectively. Then the parameters of the PID are determined as P=0.6, I=30 and D=0.006. In this case, vibration of the platform where the VSIG is installed recorded by the 3T-CMG seismometer is set as the input acceleration and the residual displacement of the test mass is given as shown in FIG. 4.

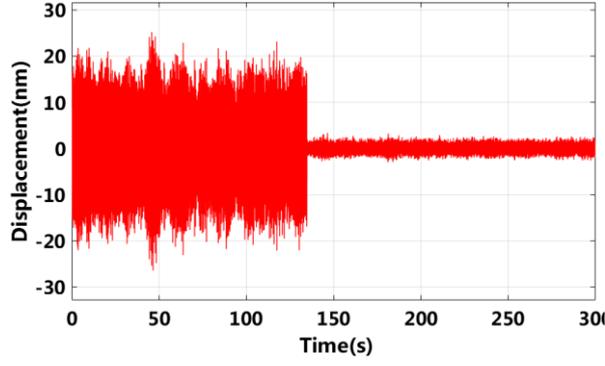

FIG. 4. displacement control simulation

Without the control, the translational displacement of the test mass is about 20nm, while reduced to 2nm when the is in closed-loop. A similar simulation of the rotational model has also been carried out.

## III. TEST RESULT AND DISCUSSION

### A.     Motion decoupling verification

Before the control, the attitude of the test mass was adjusted first to eliminate the parameter mismatch of the side-wall coils.

In the adjustment, Sinusoidal current with an amplitude of 10mA and frequency of 1Hz was injected into the translational detecting superconducting circuits in FIG. 2 through the feedback transformer on the left side. The forces produced by this sinusoidal current is the same as the one in equation (13), this force will drive the test mass to vibrate. Due to the tilt of the test mass, mismatch exists in the parameters of the coils, so both translation and rotation will be produced. Then the translation and rotation in the non-sensitive axis will couple to the vertical direction. The coupling coefficients (defined as the ratio of the two amplitude of the response signals in the non-sensitive and sensitive axes to the applied forces) between the non-sensitive and sensitive axes are directly determined by the tilt. The smaller the tilt, the lower the coupling. This characteristic is a practical criterion for adjusting.

At the beginning of the adjustment, the initial cross-coupling coefficient was measured after the sinusoidal current was injected. Then persistent current in one of the loops of the rotational detecting circuit in FIG. 2(b) was changed to introduce additional static moment to tile the test mass, and the cross-coupling coefficients were measured again. If the cross-coupling coefficients became small when compared with the initial one, the adjusting was in the right direction and the persistent current in this loop was precisely adjusted step by step to approach the optimal state that the cross-coupling coefficient was the smallest. The tilt in both the non-sensitive axes was adjusted, and the measured coefficients were shown in FIG. 5:

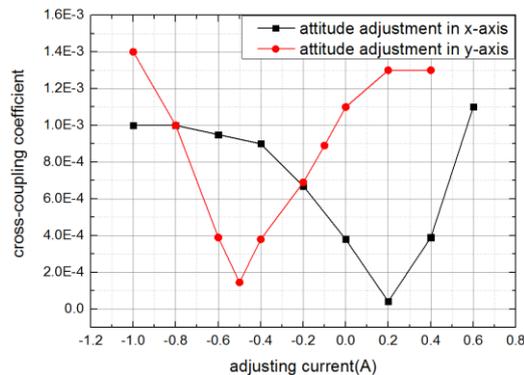

FIG. 5. Attitude adjustment of test mass by rotation circuits.

Both the optimal state had been found. Further, after the adjustment, the cross-coupling coefficients were measured again to ensure that the adjustment between the non-sensitive axes did not affect each other. The smallest cross-coupling coefficient was limited by the intrinsic magnetic coupling of the instrument, including the coupling between the superconducting coils, the transformers and the superconducting circuit loops. The intrinsic magnetic coupling coefficient was measured to be at the level of $10^{-4}$.

After the tilt was adjusted, the mismatch of the side-wall coils would be eliminated in theory. If so, when sinusoidal current was injected into the translational superconducting circuit, only translational acceleration was produced. By measuring the amplitude-frequency response of the translational detecting SQUID (SQUID1 in FIG. 2), one could find only one peak at the natural frequency of the translation in the amplitude-frequency response curve.

In order to verify it, sinusoidal current with fixed amplitude was injected into the translational superconducting circuit, the frequency of the sinusoidal current increased from 1Hz to 40Hz. The response of SQUID1 was also recorded and the amplitude-frequency response curve could be plotted, as shown in FIG. 6. The amplitude-frequency response of the rotational motion was also measured using the same method. One peak existed in the two curves respectively at the rotational natural frequency 13Hz and the translational natural frequency 19Hz. The measured result was consistent with the theoretical analysis, indicating that the mismatch of the parameters of the side-wall coils had been eliminated and the translation and rotation motion of the test mass was decoupled by the special designed superconducting circuits.

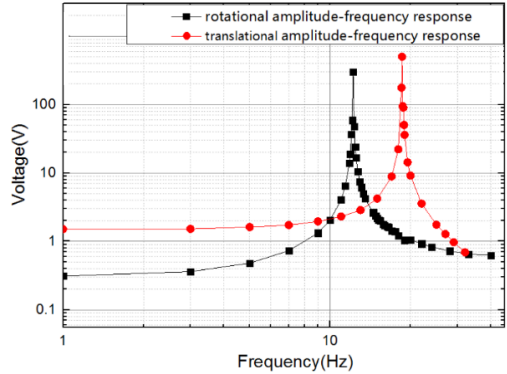

FIG. 6. The amplitude-frequency response of two kinds of superconducting circuits.

### B. non-sensitive axis displacement control

In the implementation of the non-sensitive axis displacement control, analog PID controller was used. The parameters of the PID were set according to the simulated results. The voltage-to-current converter was a highly precise resistor with 1KΩ.

Firstly, the control of the translation in one non-sensitive axis was carried out, the control result was shown in FIG. 7. Before the control, the translational displacement of the test mass was about 20nm. After control, the translational displacement was within 2nm, reduced by about one order of magnitude. In FIG. 7, there was a DC offset in the displacement before the control, the DC offset was induced because of the working principle of the SQUID [19] and did not contain any information of the displacement. Although this signal will change the attitude of test mass through the corrective action of the controller. In the real closed loop case, by increasing the sensitivity of the displacement to the voltage, the attitude effect caused by DC offset can be ignored. On the other hand, a small DC current also can be injected into the feedback transformer, adjusting the position of test mass slightly:

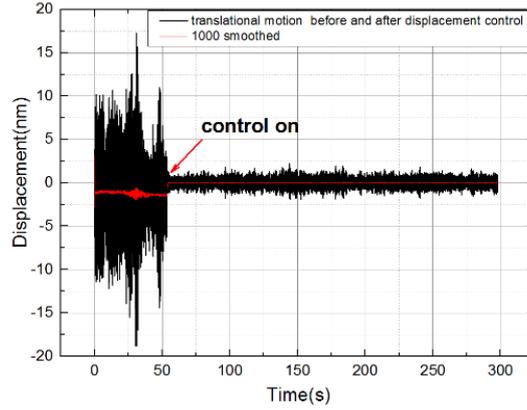

FIG. 7. Translational motion of test mass before and after control

After the control method was verified in the translational motion, the four-degrees-of-freedom (two translations and two rotations) displacement control was carried out by controlling the displacement of the test mass in the four non-sensitive axes at the same time. The final result was shown in FIG. 8. It can be seen that the four SQUID outputs are pulled back to zero after the response time of each control system and the output voltages of the four SQUIDs are suppressed, showing that the displacement of the test mass in the four non-sensitive axes has been controlled successfully.

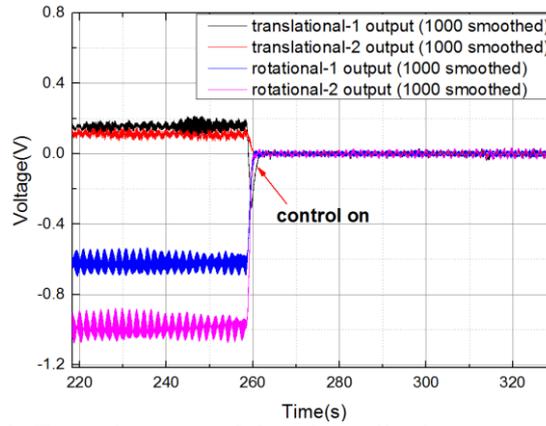

FIG. 8. Four-degrees-of-freedom displacement control

At the same time, the Power Spectral Density(PSD) of translation and rotation before and after the control is given, as shown in FIG. 9. Displacement of test mass at low frequency in closed loop is effectively suppressed compared with the one in open-loop state, showing that the equivalent stiffness of test mass in the non-sensitive axis was improved.

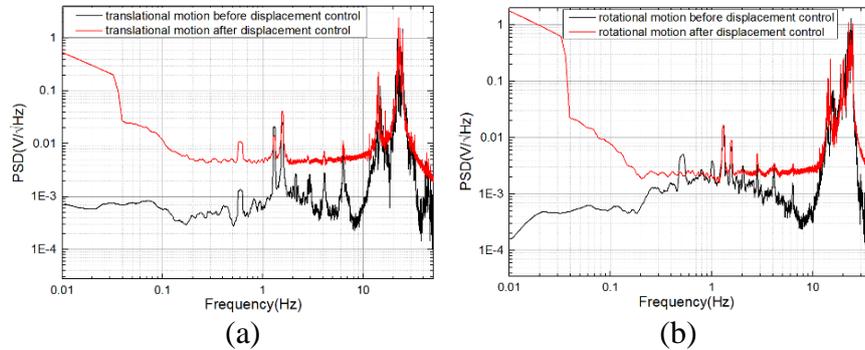

(a) (b)

FIG. 9. The PSD of translational and rotational motion before and after control: (a) translation motion (b) rotational motion

Then the noise level of SQUID output voltage of the VSGI in the sensitive axis was compared before and after the four-degrees-of-freedom displacement control, as shown in FIG. 10.

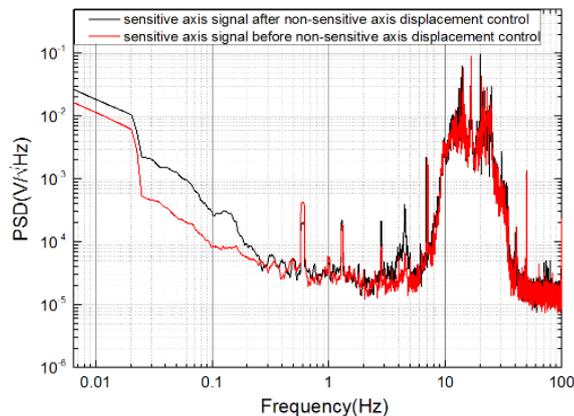

FIG. 10. Sensitive axis signal before and after non-sensitive axis displacement control

The noise level of the VSGI in the sensitive axis was reduced by about a factor of 4-6@0.03-0.1Hz, showing that the cross-coupling noise between the non-sensitive and sensitive axes has been suppressed by the non-sensitive axis control. The low-frequency noise has not yet reached background noise, and the source of 1/f may be due to a slow tilt change of the platform or a change in temperature, which requires further study. The noise level in the middle frequency band has reached the background noise of the instrument. High frequency is dominated by vertical vibration signals. Notice that there is an abnormal peak of 0.6Hz in the PSD. This may be a mode of motion caused by the cylindricity error of the test mass, as there is no constraint on the rotation of test mass about z axis. Other peaks like 1.3Hz and 2.8Hz comes from the three-wire pendulum suspension system of the VSGI, these peaks are the natural frequencies of the oscillation modes of the suspension system.

The non-sensitive axis feedback control has laid the technical foundation for suppressing the cross-coupling noise of the full-maglev VSGI, especially for the superconducting gravity gradiometer when working in the dynamic environment [9-11].

## IV. CONCLUSION

The non-sensitive axis feedback control of the test mass in full-maglev VSGI has been proposed and verified. The control system utilizes specially designed superconducting circuits to decouple the four-degrees-of-freedom motion system of the test mass into the combination of four independent one-degree-of-freedom systems, thus the PID controller can be used in the system. Test result shows that the motion of the test mass was decoupled after a precise attitude adjustment of the test mass and then the displacement of the test mass was controlled within 2nm in the laboratory by a reduction factor of one order of magnitude. Accordingly, the noise level of the VSGI in the sensitive was also reduced. The verification of this technology is of great significance to the suppression of cross-coupling noise, which lays a technical foundation for the stable operation of the instrument in dynamic environment.

## ACKNOWLEDGMENTS

This work is supported by the National Key R&D Program of China (Grant No. 2018YFC1503701) and the National Natural Science Foundation of China (Grant No.41904111).

[1] J. M. Lumly, J. P. White, G. Barnes, D. Huang, and H. J. Paik, A superconducting gravity gradiometer tool for exploration, In *Proceedings of the ASEG-PESA Airborne Gravity 2004 Workshop, Sydney, Australia , 2004* (2004).

[2] C. E. Griggs, M. V. Moody, R. S. Norton, H. J. Paik and K. Venkateswara, Sensitive Superconducting Gravity Gradiometer Constructed with Levitated Test Masses, Phys. Rev. Appl. 8, 064024 (2017).

[3] D. Ma, X.-K Liu, M. Zhang, N. Zhang, L. Chen and X.-D Liu, Wide-band Vertical Superconducting accelerometer for Simultaneous Observations of Temporal Gravity and Ambient Seismic Noise, Phys. Rev. Appl. 12, 044050 (2019).

[4] J. Hinderer and D. Crossley, Scientific achievements from the first phase (1997–2003) of the global geodynamics project using a worldwide network of superconducting gravimeters, J. Geodyn. **38**, 237 (2004).

[5] M. V. Moody, H. J. Paik, and E. R. Canavan, Three-axis superconducting gravity gradiometer for sensitive gravity experiments, Rev. Sci. Instrum. **73**, 3957 (2002).

[6] M. V. Moody, A superconducting gravity gradiometer for measurements from a moving vehicle, Rev. Sci. Instrum. 82, 094501(2011).

[7] H. A. Chan and H. J. Paik, Superconducting gravity gradiometer for sensitive gravity measurements. I. Theory, Phys. Rev. D **35**, 3551 (1987).

[8] H. A. Chan, M. V. Moody, and H. J. Paik, Superconducting gravity gradiometer for sensitive gravity measurements. II. Experiment, Phys. Rev. D **35,** 3572 (1987).

[9] C. Jekeli, Airborne gradiometer error analysis, Surv. Geophys, 27, 257–275 (2006).

[10] D. DiFrancesco, A. Grierson, D. Kaputa and T. Meyer, Gravity gradiometer systems: Advances and challenges, Geophys Prospect, 57, 615–623 (2009).

[11] M. I. Evstifeev, The State of the Art in the Development of Onboard Gravity Gradiometers, Gyroscopy Navig, 8, 68–79 (2017).

[12] F. Han and, et al. Performance of a Sensitive Micromachined Accelerometer With an Electrostatically Suspended Proof Mass[J], IEEE SENS J, 15(1):209-217 (2015).

[13] A Visioli, Practical PID Control[J], *Springer London* (2006).

[14] Yu C C, Autotuning of PID Controllers[M], *Springer London* (2006).

[15] L. Wang, D. Chen and, et al. Method for Translation and Rotation Decoupling of Test Mass in Full-Maglev Vertical Superconducting Gravity Instruments[J], Sensors, 20(19):5527 (2020).

[16] R. Torii, SQUID position sensor development, Class. Quantum Grav. **13**, 11A (1996).

[17] W. Vodel, S. Nietzsche, H. Koch, and G. J. V. Zameck, DC SQUID-based position detector for gravitational experiments, Appl. Supercond. **6**, 767 (1999).

[18] R. Scharnweber and J. M. Lumley, Characterization of a dc squid based accelerometer circuit for a superconducting gravity gradiometer, Supercond. Sci. Tech. **12**, 813 (1999).

[19] R. L. Fagaly, Superconducting quantum interference device instruments and applications, Rev. Sci. Instrum. **77**, 101101 (2006).

[20] X.-K Liu, D. Ma, L. Chen, and X.-D Liu, Tuning the stiffness balance using characteristic frequencies as a criterion for a superconducting gravity gradiometer, Sensors **18**, 1 (2018).